\documentclass[%
 aip,
 rsi,
 amsmath,amssymb,
 reprint,
]{revtex4-1}

\usepackage{graphicx}
\usepackage{dcolumn}
\usepackage{bm}
\usepackage[utf8]{inputenc}
\usepackage[T1]{fontenc}
\usepackage{mathptmx}
\usepackage{etoolbox}
\usepackage{xcolor}
\usepackage[normalem]{ulem}
\usepackage{hyperref}

\definecolor{revisionpurple}{RGB}{112,48,160}

\makeatletter
\def\@email#1#2{%
 \endgroup
 \patchcmd{\titleblock@produce}
  {\frontmatter@RRAPformat}
  {\frontmatter@RRAPformat{\produce@RRAP{*#1\href{mailto:#2}{#2}}}\frontmatter@RRAPformat}
  {}{}
}%
\makeatother

\newcommand{\ltx}{LTX-$\beta$}
\newcommand{\lya}{Lyman-$\alpha$}
\newcommand{\zeff}{Z_{\mathrm{eff}}}
\ifdefined\markedversion
  \newcommand{\add}[1]{\textcolor{revisionpurple}{#1}}
  \newcommand{\del}[1]{\textcolor{red}{\sout{#1}}}
\else
  \newcommand{\add}[1]{#1}
  \newcommand{\del}[1]{}
\fi

\begin{document}

\title[Photodiode based multi-modal diagnostic for low-energy neutral beam injection in the \texorpdfstring{LTX-$\beta$}{LTX-beta} spherical tokamak]%
{Photodiode based multi-modal diagnostic for low-energy neutral beam injection in the \texorpdfstring{LTX-$\beta$}{LTX-beta} spherical tokamak}

\author{A. Maan}
\author{T. Le}
\author{D.P. Boyle}
\author{R. Majeski}
\author{S. Banerjee}
\author{G.J. Wilkie}
\author{M. Lampert}
\author{C. L\'opez P\'erez}
\author{R. Shousha}
\affiliation{Princeton Plasma Physics Laboratory, Princeton, New Jersey 08540, USA}
\author{W. Capecchi}
\author{H. Gajani}
\affiliation{Physics, University of Wisconsin-Madison, Madison, Wisconsin 53706, USA}
\email{amaan@pppl.gov}

\date{\today}

\begin{abstract}
We present a compact photodiode-based diagnostic array developed to study low-energy neutral beam injection in the \add{Lithium Tokamak Experiment} (\ltx{}) spherical tokamak.  The in-vacuum diagnostic combines filtered soft-x-ray (SXR), narrowband \lya{}, and unfiltered absolute extreme ultra-violet (AXUV) photodiode rows with partly overlapping, nearly coincident tangential views of the plasma, including the neutral beam path.  This geometry provides simultaneous sensitivity to beam-induced SXR emission; neutral-hydrogen line radiation associated with recycling, fast neutrals and fueling; and broadband emission that can include direct neutral impacts from fast-ion charge-exchange losses.  Initial measurements from 12--20 keV hydrogen beam operation show beam-synchronous detector responses in all three modalities.  \del{The unfiltered AXUV signals exhibit millisecond-scale rise and fall times that are much slower than the detector response, that vary across sightlines, and depend on lithium-conditioning history.}\add{The unfiltered AXUV signals exhibit millisecond-scale rise and fall times that vary across sightlines and depend on lithium-conditioning history.}  Comparison with classical slowing-down time estimates indicates that charge exchange with background neutrals contributes appreciably to the measured decay. The diagnostic can potentially be used to constrain a forward model to estimate the time-resolved balance of beam heating and fueling for small tokamaks. 
\end{abstract}

\maketitle

\section{Introduction}

Neutral beam injection (NBI) can be an attractive actuator for tokamaks because it can supply auxiliary heating, non-inductive current drive, diagnostic stimulus, and core particle fueling with a single system.  In small devices, low-energy NBI must balance penetration, shine-through, prompt orbit loss, and collisional slowing down.  On \ltx{}, the beam ion source generates hydrogen beam energies of 12--20 keV.  The resulting fast neutrals must be captured without excessive shine-through, captured  fast ions must remain confined on orbits comparable to or smaller than the minor radius, and slow down on background electrons and ions before they are lost to charge exchange with background neutrals.  Initial \ltx{} NBI experiments and modeling found prompt loss of nearly all beam ions in some operating regimes; higher-current operation and optimized beam geometry were predicted to recover substantial coupled beam power \cite{Hughes_2021,Capecchi_2021}.  Since energetic-particle confinement in small tokamaks is strongly coupled to edge recycling and neutral-source physics \cite{McCracken_1979,stangeby,TSUJI1991311}, measurements that simultaneously diagnose beam deposition, edge neutral emission, and radiated power are needed to interpret low-energy NBI.

\begin{table*}[t!]
\caption{\label{tab:detector-geometry}Summary of the photodiode detector geometry. The etendue range is the channel-by-channel estimate $G=A_{\mathrm{diode}}A_{\mathrm{slit}}\cos^2\theta/d^2$, where $d$ is the slit-to-diode distance and $\theta$ is the incidence angle relative to the slit and detector normals. \add{A$_{\mathrm{diode}}$ and A$_{\mathrm{slit}}$ are the effective areas of the photodiode and slit, respectively}. $L_{Beam}$ is the projected distance along the neutral beam centerline from the beam entry point to the sightline intersection; \add{$L_{AXUV}$ is estimated path length of the sightline into the torus, fig~\ref{fig:toroidal-sightlines} has a pictorial depiction}. \add{The elevation is the vertical distance from the toroidal midplane to the center of the slit. Details of the sightline calculations, including scripts to reproduce them, are included in the data repository \cite{tor_rsi_data}.}}
\begin{ruledtabular}
\begin{tabular}{l c c c c c}
Diagnostic & $G$ & $r_{\tan}$ & $L_{Beam}$ & $L_{AXUV}$ & \del{Height}\add{Elevation} \\
& (mm$^2$ sr) & (mm) & (mm) & (mm) & (mm) \\
\hline
AXUV bolometer & $1.15$--$1.19\times10^{-4}$ & 205--425 & 525--862 & 1023--1277 & -30.5 \\
\lya{}, low-field side & $1.67$--$1.75\times10^{-3}$ & 287--450 & 661--894 & 1005--1217 & -15.2 \\
\lya{}, high-field side & $1.48$--$1.58\times10^{-3}$ & 156--316 & 433--710 & 1195--1308 & 0.0 \\
SXR, high-field side & $1.11$--$1.18\times10^{-3}$ & 156--317 & 434--711 & 1185--1308 & 15.2 \\
SXR, low-field side & $1.18$--$1.22\times10^{-3}$ & 289--422 & 664--858 & 1053--1217 & 30.5 \\
\end{tabular}
\end{ruledtabular}
\end{table*}

\begin{figure}[b!]
\includegraphics[width=0.55\columnwidth]{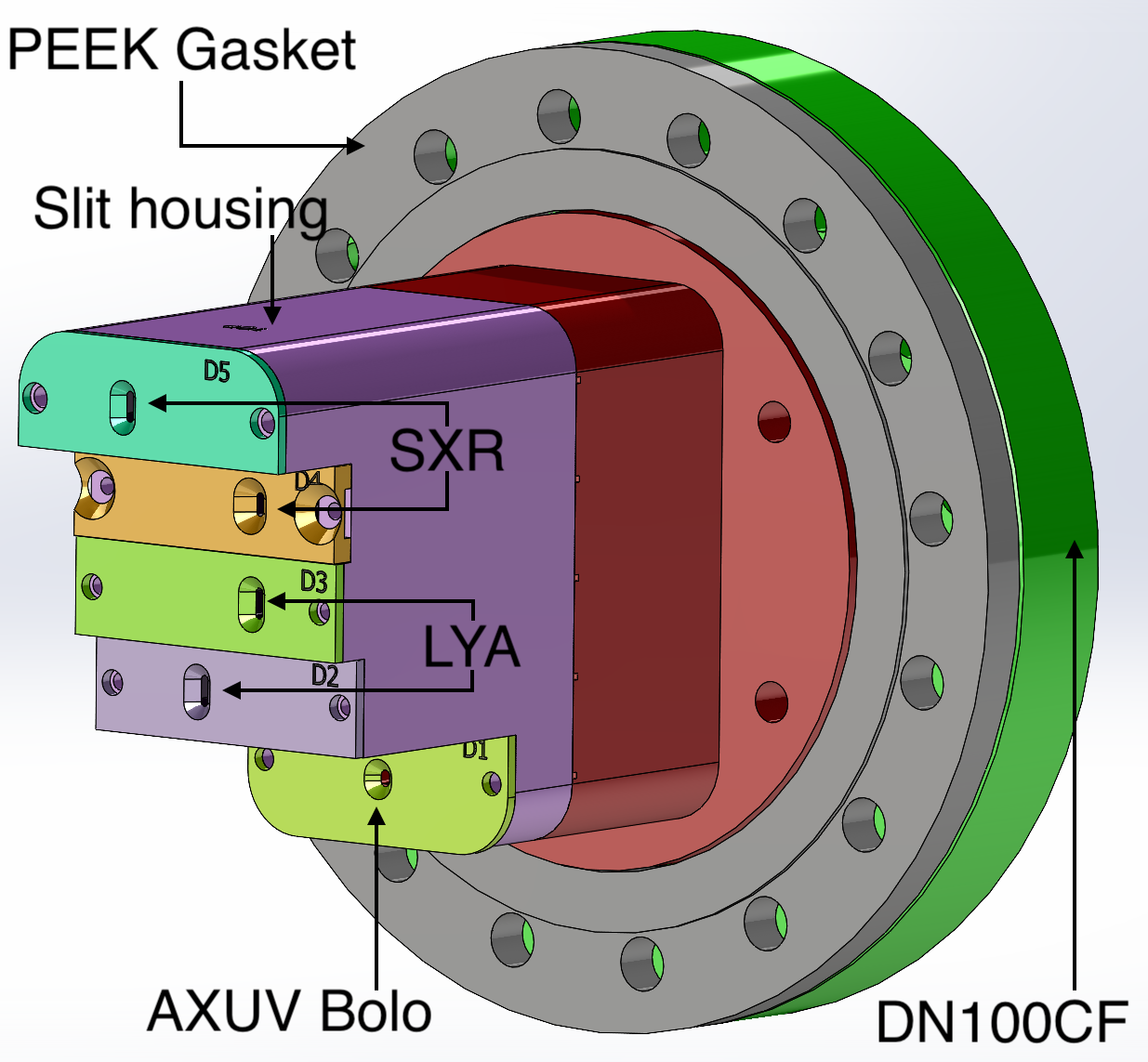}
\caption{\label{fig:tor-array}Vacuum-side toroidal photodiode array assembly model showing the diagnostic housing and front-slit/filter hardware mounted to the flange. Position of the slits is annotated for the three modalities.}
\end{figure}

\ltx{} is a low-aspect-ratio tokamak designed to study plasma operation with a low-recycling lithium boundary.  Lithium retains hydrogen isotopes and can reduce recycling from plasma-facing components \cite{Baldwin_2002,Kaita_2019}.  Low recycling has long been predicted to raise edge temperature, flatten electron-temperature profiles, and weaken temperature-gradient-driven transport \cite{crashninicove,isomak}.  Previously, LTX demonstrated hot-edge, nearly flat electron temperature profiles with lithium-coated walls \cite{boyle-prl,majeski_pop}. Later, \ltx{} experiments directly connected the hot edge regime to measurements of reduced recyling and improved neutral and plasma density control with increasing lithium coatings \cite{MAAN2023_NME,Maan_2024_recycling}.  \ltx{} operation extended this regime with both ohmic and neutral-beam heating, with edge temperatures approaching core values and confinement exceeding conventional ohmic and H-mode scalings \cite{Boyle_NF_2023}.  Because edge gas puffing can cool the hot, sparse edge, NBI core fueling is a natural tool for sustained low-recycling operation. \add{These considerations motivate a diagnostic capable of measuring the spatial and temporal evolution of beam--plasma interactions in these unique plasmas, including beam heating and fueling, the fast-ion population, and changes in plasma--material interactions.}


\begin{figure}[htbp]
\centering
\includegraphics[width=\columnwidth]{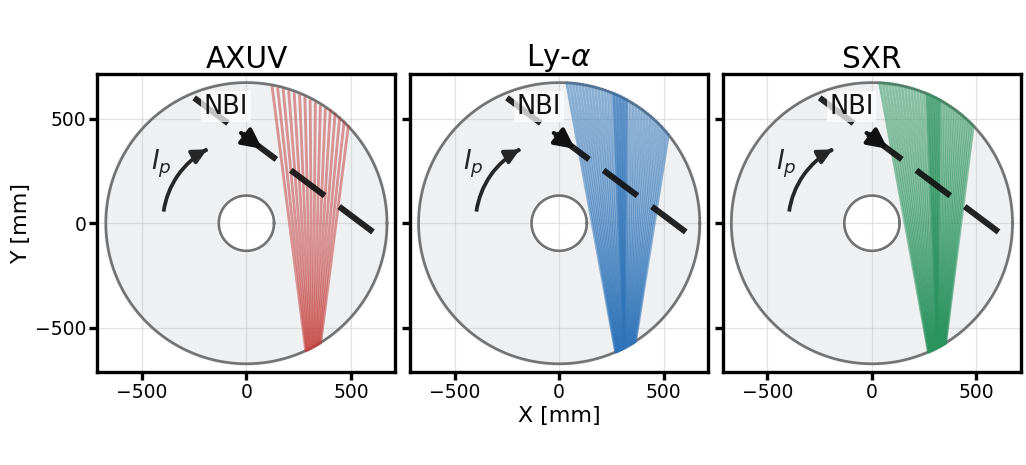}
\caption{\label{fig:toroidal-sightlines}Toroidal projection of the photodiode sightline centerlines \add{tabulated in Table~\ref{tab:detector-geometry} as $L_{AXUV}$} and the NBI centerline, split by modality. The circular $I_p$ arrow shows the plasma-current direction; the neutral beam is co-injected in \ltx, here shown at r$_{tan} \sim $ 33 cm.}
\end{figure}

\renewcommand{\topfraction}{0.95}
\renewcommand{\bottomfraction}{0.95}
\renewcommand{\textfraction}{0.02}
\renewcommand{\floatpagefraction}{0.8}
\setcounter{topnumber}{4}
\setcounter{bottomnumber}{4}
\setcounter{totalnumber}{8}

This paper presents a compact multi-modal photodiode array for low-energy NBI studies on \ltx{} (Fig.~\ref{fig:tor-array}). \add{The diagnostic combines filtered SXR measurements of continuum and impurity-line emission sensitive to \(T_e\) and \(Z_{\mathrm{eff}}\), narrowband \lya{} measurements of neutral-hydrogen emission associated with recycling and fueling, and broadband AXUV measurements sensitive to radiated power and possible direct impacts of fast-neutral particles on the detector.} \del{The diagnostic combines soft x-ray, \lya{}, and broadband AXUV measurements to separate beam-heated radiation, neutral-hydrogen emission associated with recycling and fueling, and fast-neutral or bolometric response.} AXUV photodiodes have been used as fast neutral-particle detectors in the past \cite{Clary_2012,Veshchev_2006,Zhu_2012,Liu_2016}.  Similar photodiode approaches have also been used for fast bolometry and \lya{} measurements in tokamaks \cite{Degeling_2004,lama_cal}; here the implementation is tailored to limited port access, low beam energy, and changing lithium-wall conditions. Section II describes the diagnostic design and implementation. Section III demonstrates the response of the various diagnostic arrays to NBI-fueled \ltx{} discharges, and explains how the measurements can be used to investigate the unique physics of \ltx{}. Section IV discusses observations of variations in the decay time of the unfiltered AXUV signal following termination of the NBI and connects them to Li wall conditions. Section V concludes with a summary and possibilities to constrain future modeling.

\del{The diagnostic is particularly novel in application to the unique environment of \ltx{}, where it is important to understanding key physics.} \add{The diagnostic is particularly well suited to the unique environment of \ltx{}, where it enables studies of key low-recycling physics, including lithium-aided core fueling, charge-exchange effects under reduced-neutral conditions, and impurity effects associated with lithium plasma--material interactions \cite{Krstic_2023_LiPMI}.} Data from the diagnostic was recently used to simulatensouly diagnose island width, location and rotating frequency by forward modelling the SXR intensity. Magnetic islands in \ltx{} are related to tearing mode activity with a known recycling and edge neutral density dependence \cite{tosh,Banerjee_2024}. 

\section{Design, viewing geometry and instrumentation}

The diagnostic uses windowless AXUV20ELM silicon photodiode arrays \cite{OptoDiodeAXUVSeries}, avoiding entrance-window attenuation of VUV, EUV, and soft x-ray photons.  Each element has a $0.75~\mathrm{mm}\times4.1~\mathrm{mm}$ active area, is expected to have 100\% internal quantum efficiency, and has photon-response curves spanning the EUV-to-UV and UV-to-near-infrared ranges.  In the soft x-ray band, absorbed photon energy produces electron-hole pairs in silicon, giving an approximately flat energy-to-current responsivity of $\sim$ $0.26~\mathrm{A/W}$. 

The installed head contains five vertically stacked rows: two 20-channel soft-x-ray (SXR) arrays, two 20-channel \lya{} arrays, and one 16-channel unfiltered AXUV array.  The SXR and \lya{} rows form high-field-side and low-field-side groups with nearly overlapping tangential views (Fig.~\ref{fig:toroidal-sightlines}). The diagnostic was initially tested on \ltx{} installed on a similar tangential port on the other side of the machine\cite{tosh}, essentially the same view translated from x~300 mm to x~-300mm. The unfiltered AXUV row uses a commercial $50~\mu\mathrm{m}$ air slit, while the SXR and \lya{} views are defined by precision-machined apertures bolted to the slit housing.  The slit/filter hardware is carried by the front half of a two-piece 316 stainless steel enclosure; the rear half mounts to the vacuum-side flange and houses the diode arrays.  Pump-out trenches between the halves prevent trapped gas volumes while preserving the light-tight enclosure.

The SXR channels use $5~\mu\mathrm{m}$ beryllium filters bonded over the machined apertures.  For response modeling, CXRO/Henke Be transmission for density $1.848~\mathrm{g\,cm^{-3}}$ and thickness $5~\mu\mathrm{m}$ was tabulated from 100 to 4000 eV and \del{convolved with}\add{multiplied pointwise by} AXUV responsivity \cite{Henke1993,CXROFilterTransmission}.  \add{This pointwise product gives the effective spectral response of the Be-filtered SXR channels used to interpret the passband below.}  The passband is effectively closed below a few hundred eV (blocking \add{all} Li emission), rises through 0.4--1 keV, and approaches the unfiltered AXUV response at several keV.  Without significant C or O line contamination, the filtered SXR brightness is primarily line-integrated bremsstrahlung and can constrain $\zeff{}$ with electron density and temperature profiles.  Otherwise it should be interpreted as an effective SXR emissivity unless impurity radiation is modeled explicitly.

The \lya{} rows use 0.5 in. diameter Acton Optics VUV/UV filters centered at $122~\mathrm{nm}$ (with uncertainty $\pm2.5~\mathrm{nm}$), close to the hydrogen \lya{} wavelength of $121.6~\mathrm{nm}$.  The VUV MgF$_2$ filters have peak transmission of $\sim$ 15\% and full width at half maximum $<$ 10 nm.  Mounted in front of the machined \lya{} apertures, they select neutral-hydrogen line emission with geometry matched to the SXR rows \add{as seen in fig~\ref{fig:toroidal-sightlines}}.

The diode arrays mount on a 6 in. ConFlat (DN100CF) flange with DB25 electrical feedthroughs and low-outgassing, gold-electroplated adapter PCBs.  A light-tight stainless steel enclosure surrounds the arrays; the AXUV slit and SXR Be filters are bonded with VACSEAL, and Acktar Lambertian Black foil suppresses internal reflections.  The diagnostic is mounted behind a pneumatically actuated gate valve to allow isolation for service and to protect the diagnostic during Li evaporations, glow discharge cleaning, and Li re-evaporation from hot surfaces. PEEK and Vespel  mounting hardware electrically isolate the diagnostic body, while air-side transimpedance and differential amplifiers condition the photocurrents for 250 kHz D-TACQ digitization.

A simple signal-to-noise estimate used 41 shots across all 96 channels.  For each channel, the baseline was the pre-discharge interval $t=0.369$--$0.422~\mathrm{s}$.  The rms fluctuation in this window, $\sigma_n$, was compared with the 99th percentile of $|V-\bar{V}_{\mathrm{base}}|$ after the baseline window to define $\mathrm{SNR}_{99}=V_{99}/\sigma_n$.  Median row $\mathrm{SNR}_{99}$ values were approximately 890 for the AXUV bolometer, 170 for low-field-side \lya{}, 96 for high-field-side \lya{}, 55 for high-field-side SXR, and 94 for low-field-side SXR; the lowest usable channels were above $\mathrm{SNR}_{99}\simeq40$.

\section{Initial data from beam fueled discharges}

Initial NBI coupling experiments on \ltx{} were limited by prompt fast-ion loss: with the beam at lower tangency radius, fast ions were lost to plasma-facing components before transferring substantial heat or enhancing density \cite{Hughes_2021,Capecchi_2021}.  Three-dimensional orbit calculations indicated that a higher tangency radius would reduce these losses, so the beam was re-aimed to $r_{\tan}=33~\mathrm{cm}$ for the final \ltx{} campaign \cite{Zakharov2026LTXBetaReport}.  The discharges discussed here used the re-aimed configuration.

\begin{figure}[htbp]
\centering
\includegraphics[width=\columnwidth]{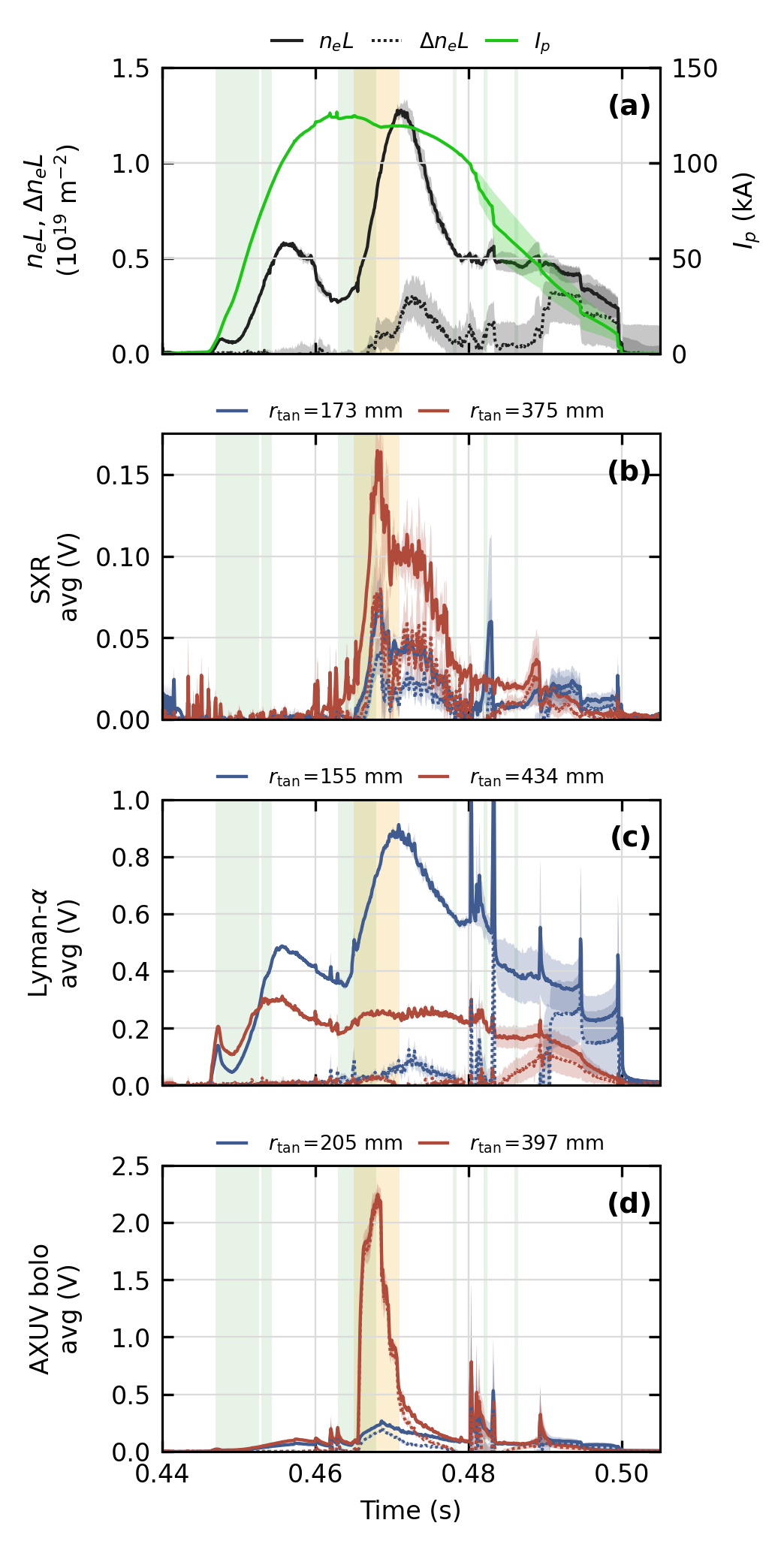}
\caption{\label{fig:beam-shot}Ensemble-averaged beam-fueling discharge.  The ensemble contains five beam shots and five no-beam comparison shots.  Panel (a) shows beam-shot averaged $I_p$ and $n_eL$ together with no-beam-subtracted $\Delta n_eL$. $I_p$ sees no measurable difference between the beam and no-beam shots.  Panels (b)--(d) show beam-shot averaged SXR, \lya{}, and AXUV bolometer signals for selected high-field-side and low-field-side channels; dashed traces are beam-minus-no-beam differences.  Shaded bands are SEMs, propagated in quadrature for dashed differences.  Yellow shading marks the 465--471 ms NBI pulse from the first beam shot, and green shading marks the high-field-side gas-puff intervals from the same shot.  Diode legends list the selected sightline $r_{\tan}$ values.}
\end{figure}

Figure~\ref{fig:beam-shot} shows a representative ensemble from later in the campaign.  The NBI pulse was programmed from 465 to 471 ms; the beam shot had average beam current $30.5~\mathrm{A}$ and voltage $16.7~\mathrm{kV}$.  The excess density associated with the beam (calculated by subtracting off the density of an identical no-beam shot) reaches $\Delta n_eL\simeq2.4\times10^{18}~\mathrm{m}^{-2}$ over 471--476 ms, about 23\% of the beam-shot line density.  Since the beam current alone cannot supply this charge, the observed fueling is likely a combination of direct beam fueling and beam-driven recycling. \add{The additional particle source may arise from beam-enhanced recycling, in which charge-exchange products from captured fast ions reach plasma-facing surfaces and modify the wall particle inventory and recycling source \cite{Hogan1997,Bagryansky1999}.}

All three modalities show clear detector response to the beam in Figure~\ref{fig:beam-shot}, which shows the \add{detector response} \del{full signals} during the NBI shots as solid lines, while the dashed lines show the excess signal relative to identical no-beam baseline shots. SXR channels show beam-correlated excess signals (dashed lines) of roughly half the beam-shot SXR peak (solid lines) near the pulse.  The \lya{} excess (dashed) is relatively smaller; averaged over 471--480 ms, it is only a few percent of the high-field-side beam-shot signal and near zero on the selected low-field-side channel after the beam turns off.  \add{The green bands in Fig.~\ref{fig:beam-shot}(c) identify the high field side gas puff intervals and help distinguish the localized fueling response from the beam-correlated \lya{} signal. The puff response is strongest on the inboard channels and decreases toward the outboard array, whereas the beam minus no-beam excess is mostly uniform across all channels during beam injection and persists after NBI turnoff on channels closer to the high field side.}  The AXUV bolometer shows the brightest beam-synchronous excess (dashed) feature, especially on the selected low-field-side chord, where the no-beam-subtracted (dashed) component accounts for nearly the full beam-time (solid line) peak.

The SXR response is not a direct measurement of deposited beam power. For continuum-dominated emission, the Be-filtered signal scales approximately as a line integral of $n_e^2\zeff{}$ multiplied by a temperature response set by Be transmission and AXUV responsivity.  Because the filter transmits mainly above a few hundred eV, this temperature dependence is strongly nonlinear at \ltx{} temperatures.  Previous LTX lithium-wall discharges provide a useful constraint: core impurity levels were low, with $\zeff{}\simeq1.2$ and an estimated lithium contribution below 0.1 to $\zeff{}$ in flat-temperature plasmas \cite{majeski_pop}.  Surface-science measurements show that lithium evaporation and oxidation affect impurity sources, including lithium-oxide formation and C/O contamination \cite{Lucia_2015_surface,Maan_PPCF_2020,Maan_IEEE_2020}.  These measurements support treating Li, C, and O as the principal impurity species in \ltx{} low-recycling plasmas \cite{majeski_pop,Lucia_2015_surface,Maan_PPCF_2020}. \add{Quantitative $\zeff{}$ inference therefore requires an independent measure of core impurities, primarily C and O, to separate the SXR continuum from line radiation.}

Lithium line radiation is strongly suppressed by the $5~\mu\mathrm{m}$ Be filter, whereas C V-VI and O VII-VIII line radiation falls closer to the rising passband.  Small C or O concentrations can therefore contribute disproportionately to filtered SXR signal.  \add{If omitted, this component can make a beam-correlated SXR increase appear as larger continuum $\zeff{}$, stronger electron-temperature rise, or larger coupled beam power and is a key limitation of the diagnostic. This would not be the case in a tokamak with higher core electron temepratures where the use of a thicker Be foil would strongly supress C and O line radiation as well as Li}.  The \ltx{} HAL charge-exchange recombination spectroscopy diagnostic can constrain this ambiguity where Li III/O II ratios are available, given suitable atomic physics and profile information \cite{Elliott_2018_CHERS}. 

The \lya{} signal can be calculated by $S_{\mathrm{Ly}\alpha}\propto\int n_e n_H \mathrm{PEC}_{\mathrm{Ly}\alpha}(n_e,T_e)\,dl$, and is thus sensitive to the beam-driven neutral source and its spatial distribution. Previous \ltx{} recycling estimates used \lya{} measurements to constrain DEGAS2 neutral-transport calculations, but the model had difficulty matching edge poloidal channels because SOL $n_e$, $T_e$, and radial decay lengths near limiting surfaces were not sufficiently constrained \cite{Maan_2024_recycling,DEGAS2}.  The present toroidal array grazes the high-field-side limiter with useful $r_{\tan}$ coverage, adding direct sensitivity to the wall-recycling source and therefore better constrain Scarpe-Off Layer (SOL) radial decay lengths.  With a constrained SOL model, these data can infer background neutral distributions, estimate beam-ion charge-exchange loss rates, and further constrain neutral-beam/SXR forward modeling. \add{Without a forward model, the \lya{} signal can be interpreted as a qualitative measure of recycling and fueling, with the high-field-side channels more sensitive to recycling from the limiter and the low-field-side channels more sensitive to fueling from the beam, fig~\ref{fig:beam-shot}.}


\section{Discussion: AXUV bolometer decay times and lithium conditioning}

\del{The bright beam-correlated AXUV signal can be interpreted as a fast-neutral signal from confined beam ions that re-neutralize by charge exchange and leave the plasma.}\add{The bright beam-correlated AXUV signal may include detector current produced by fast neutral particles incident directly on the diode. These particles can originate when confined beam ions re-neutralize by charge exchange and escape the plasma.} AXUV photodiodes have been used this way as compact fast-neutral analyzers, because escaping neutrals deposit kinetic energy in the diode and produce a signal proportional to neutral particle flux and detector energy response \cite{Clary_2012,Veshchev_2006,Zhu_2012,Liu_2016}.  AXUV responsivity alone does not exclude \add{a photocurrent contribution from} visible line radiation.  \del{However, the small beam-correlated \lya{} response, the strong low-field-side spatial localization of the AXUV excess, and the $r_{\tan}$-dependent millisecond decay make a purely prompt photon explanation unlikely.}\add{The AXUV excess exhibits an $r_{\tan}$-dependent millisecond-scale decay after beam turnoff, as quantified below. This delayed response is inconsistent with a signal arising solely from photon emission produced promptly by direct beam excitation, although slower plasma-radiation contributions cannot be excluded.}  For now we treat the AXUV signal as an effective fast-neutral/bolometric channel until a forward model separates photon and neutral-particle contributions.

\begin{figure}[htbp]
    \centering
    \includegraphics[width=0.95\columnwidth]{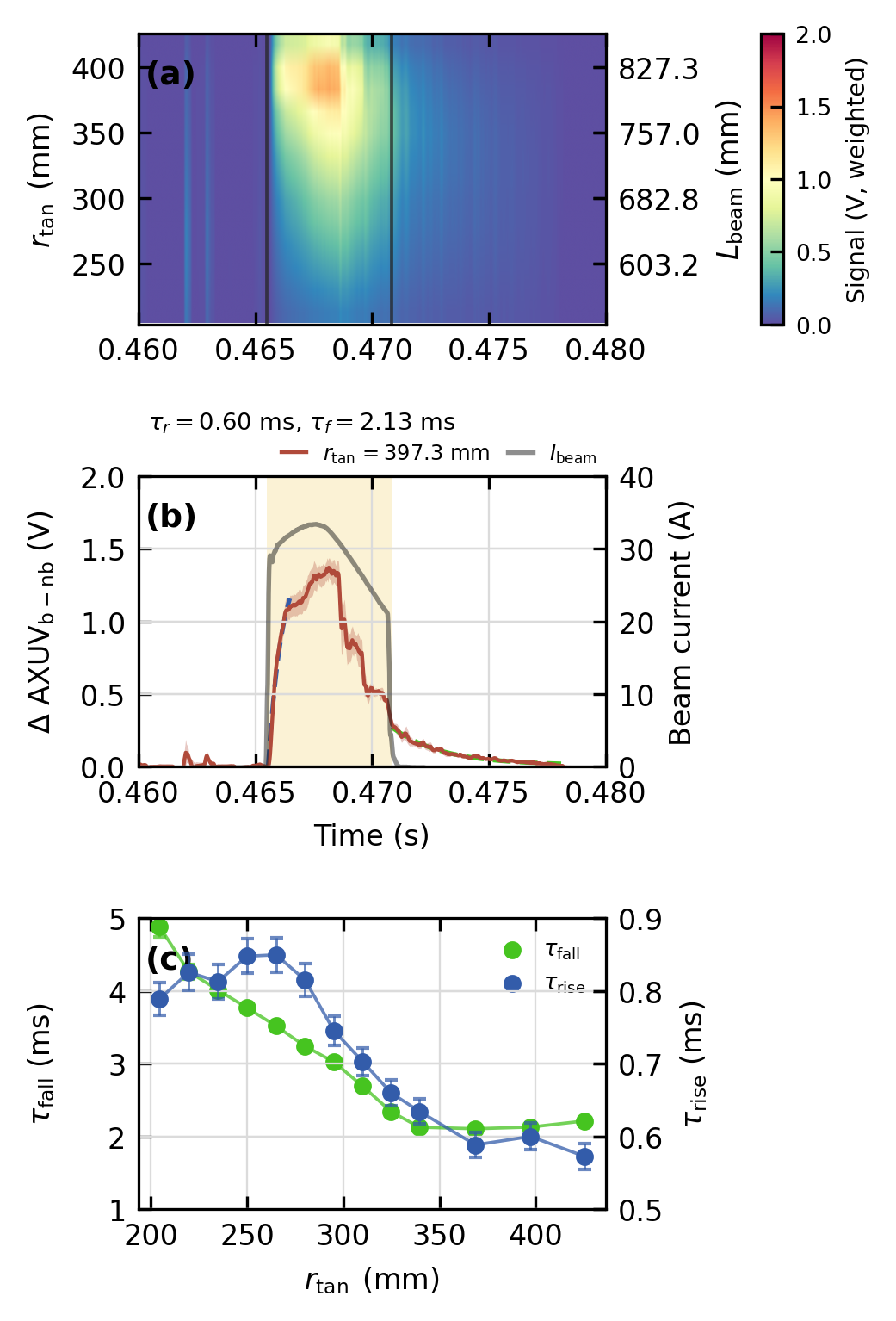}
    \caption{Beam-minus-no-beam AXUV response for a highly lithium-conditioned ensemble (a) AXUV signal versus tangency radius, with the beam path length shown on the right axis.  (b) Brightest beam-minus-no-beam trace with propagated uncertainty band, exponential rise and fall fits, and the beam-current trace from the ensemble median beam shot.  (c) Fitted fall and rise times for all channels.}
    \label{fig:axuv-tau-dec11}
\end{figure}

\add{As shown in Fig.~\ref{fig:axuv-tau-dec11}(a,b), the beam-minus-no-beam AXUV signal rises over a finite interval after NBI turn-on and decays after NBI turn-off, with both transients varying across sightlines}. Figure~\ref{fig:axuv-tau-dec11} shows beam-minus-no-beam response from later in the campaign.  Fitted rise times are $0.57$--$0.85~\mathrm{ms}$ and fall times are $2.1$--$4.9~\mathrm{ms}$, with the shortest decays on high-$r_{\tan}$ chords.  This spatial dependence suggests that the post-beam AXUV decay is controlled by the local fast-ion loss and slowing-down balance sampled by each sightline, not a single electronics or thermal time constant.

As a first estimate, the observed fall time can be treated as the combined loss rate from charge exchange and collisional slowing down, with prompt orbit loss assumed not to dominate based on orbit calculations for beam injection in \ltx{} at r$_{tan} >$ 330 mm \cite{Zakharov2026LTXBetaReport,Capecchi_2021}.  The H$^0$+H$^+$ charge-exchange curve gives $\sigma_{\mathrm{CX}}=6.7\times10^{-16}$ and $4.5\times10^{-16}\,\mathrm{cm}^{2}$ at 12 and 20 keV, respectively, corresponding to $\sigma_{\mathrm{CX}}v=(1.02$--$0.88)\times10^{-13}\,\mathrm{m}^{3}\,\mathrm{s}^{-1}$ \cite{Barnett1990AtomicData,Zakharov2026LTXBetaReport}.  If the decay were charge-exchange dominated, $\tau_{\mathrm{fall}}=1$--$5~\mathrm{ms}$ would therefore imply $n_0\simeq2.0\times10^{15}$--$1.1\times10^{16}~\mathrm{m}^{-3}$ for the 16.7 keV beam injection energy.  For a $300~\mathrm{eV}$ core plasma and $n_e=1$--$2\times10^{19}~\mathrm{m}^{-3}$, the classical slowing-down time is approximately $6$--$18~\mathrm{ms}$, so including finite slowing down changes the inferred neutral density by order-unity factors but leaves the characteristic range near $10^{15}$--$10^{16}~\mathrm{m}^{-3}$.  Thus charge exchange must contribute appreciably to the decay, with shorter high-$r_{\tan}$ decays requiring larger local neutral density or charge-exchange rate.  These values are consistent with reported core neutral densities in low-recycling \ltx{} discharges.

Lithium conditioning history is important because inventory and oxidation affect recycling, impurity influx, density control, and plasma performance on \ltx{} \cite{Maan_PPCF_2020,MAAN2023_NME,Maan_2024_recycling}. \add{Run-day analysis after lithium evaporation also shows that plasma current, density, and particle pumping evolve on timescales consistent with oxidation of the lithium surface \cite{Maan_2026_fLITER}.}  The later in the  campaign discharges presented in Figures \ref{fig:beam-shot} and \ref{fig:axuv-tau-dec11} followed an evaporation that deposited 403 nm on average on the quartz crystal microbalances, corresponding to $2.16~\mathrm{g}$ \add{lithium deposited on the shells}\del{using the shell-area estimate}.  By this ensemble, roughly three months into the campaign, cumulative lithium evaporation on the shells was $13.0~\mathrm{g}$. \add{Earlier campaigns from \ltx{} show that conditioning is a strong function of cumulative lithium deposition, with the first evaporations after stainless-steel operation providing relatively poor recycling control and higher impurity influx than later in the campaign \cite{Maan_IEEE_2020,Maan_PPCF_2020,Maan_2024_recycling,Boyle_NME_2025,MAAN2023_NME}.}

\begin{figure}[t]
    \centering
    \includegraphics[width=0.8\columnwidth]{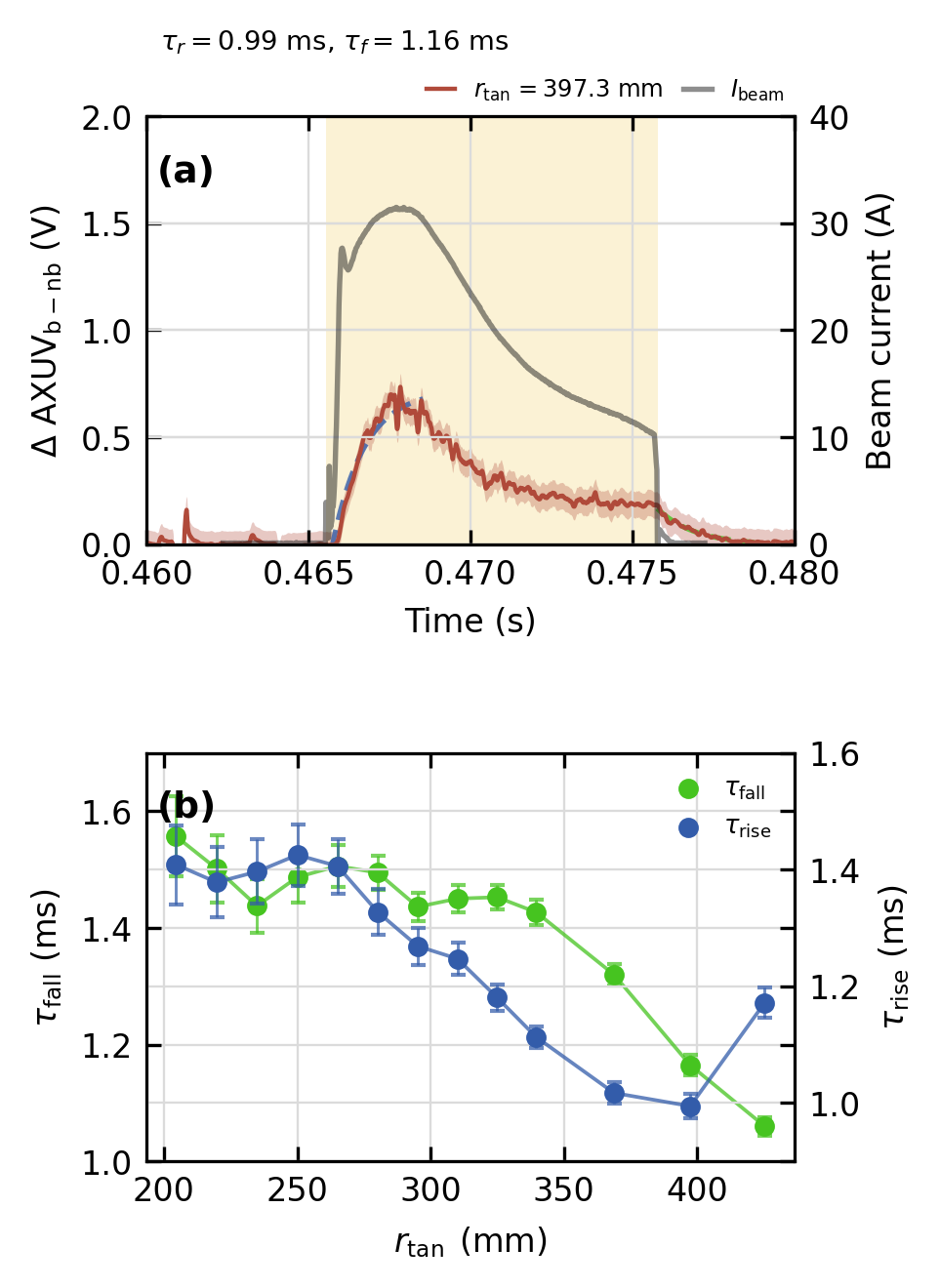}
    \caption{AXUV response from a beam-minus-no-beam shot pair after the first lithium evaporation after \ltx{} transitioned from stainless-steel wall operation.  (a) Beam-minus-no-beam trace with baseline-noise uncertainty band, exponential rise and fall fits, and the beam-current trace.  (b) Fitted fall and rise times across all channels.}
    \label{fig:axuv-tau-oct30}
\end{figure}

Shots immediately after the first lithium evaporation as \ltx{} transitioned from stainless-steel wall operation show shorter AXUV fall times compared to shots later in the campaign with greater accumulated lithium conditioning.  Figure~\ref{fig:axuv-tau-oct30} shows a comparable set of AXUV responses from the initial evaporation, with beam current approximately $20.2~\mathrm{A}$ for 16.7 keV hydrogen-beam.  The brightest channel at r$_{tan} \sim 397$ mm has $\tau_{\mathrm{fall}}\simeq1.2~\mathrm{ms}$, and all channels are approximately $1.1$--$1.6~\mathrm{ms}$, shorter than the later-campaign case in Fig.~\ref{fig:axuv-tau-dec11}.  This early-versus-late comparison is not a controlled matched-shot scan: after lithium evaporation, nominally similar shot programming does not generally reproduce the same density, temperature, gas-puff, impurity, and wall-state evolution, and the two cases also differ in beam current and ensemble size.  The trend is therefore suggestive rather than a direct isolation of lithium conditioning.  It is nevertheless robust in the available data to changes in beam injection energy and current, and it is consistent with the expectation that first evaporations after stainless-steel operation can be relatively dirty and provide poorer recycling control than a repeatedly conditioned lithium wall \cite{Maan_IEEE_2020,Maan_PPCF_2020,Maan_2024_recycling,Boyle_NME_2025,MAAN2023_NME}.  The shorter \del{early} fall times \add{earlier in the campaign} imply larger effective charge-exchange rates and therefore larger neutral density sampled by the beam-viewing chords.

Lithium also favors low-energy beam capture, the electron-loss/capture cross section for a 12--20~keV hydrogen beam interacting with fully ionized lithium is approximately $5\times10^{-15}\,\mathrm{cm}^{2}$, roughly an order of magnitude larger than resonant hydrogen charge exchange over the same range \cite{Barnett1990AtomicData,Summers2007ADAS,ADAS302ADF02,Zakharov2026LTXBetaReport}. \add{This mechanism of better beam capture with slightly higher lithium core concentration is supported by the data, $\tau_{rise}$ times of the AXUV beam minus no-beam signal is shorter when discharges were run with well conditioned walls later in the campaign compared to earlier in the campaign as seen in a comparison of fig~\ref{fig:axuv-tau-dec11}b) and  fig~\ref{fig:axuv-tau-oct30}b); where the rise time is seen to reduce by a factor of two. We hypothesize that a bigger core lithium concentration means the fast ion reservior will build faster and the corresponding fast neutral loss channel will be visible more quickly on the AXUVs}.  Since this process captures beam neutrals without re-neutralizing already-confined ions, lithium can increase the confined beam fraction without increasing charge-exchange loss.  The observed conditioning dependence could therefore be a combination of reduced recycling and increased beam capture. \add{Across the AXUV chords, well-conditioned discharges from later in the campaign exhibit shorter rise times, $0.57$--$0.85~\mathrm{ms}$, and longer fall times, $2.1$--$4.9~\mathrm{ms}$ (Fig.~\ref{fig:axuv-tau-dec11}), compared with approximately $1.0$--$1.4~\mathrm{ms}$ and $1.1$--$1.6~\mathrm{ms}$, respectively, after the first lithium evaporation (Fig.~\ref{fig:axuv-tau-oct30}). The shorter rise is consistent with enhanced beam capture and faster buildup of the fast-ion population, whereas the longer decay is consistent with reduced recycling and lower plasma neutral density.}

\section{Conclusions}

\del{The diagnostic's first integrated \ltx{} beam measurements show beam-synchronous AXUV emission with millisecond-scale rise and fall times that are longer than the detector response and spatially dependent across beam-viewing sightlines.}\add{The diagnostic's first integrated \ltx{} beam measurements show beam-synchronous AXUV signals with millisecond-scale, spatially dependent rise and fall times across the beam-viewing sightlines.} Lithium-conditioned discharges from later in the campaign, with greater cumulative lithium deposition, show fall times of a few milliseconds on high-$r_{\tan}$ chords and longer times on lower-$r_{\tan}$ chords. By contrast, discharges from the initial transition from stainless-steel-wall to lithium-conditioned-wall operation show uniformly shorter fall times near $1$--$2~\mathrm{ms}$. Compared with $6$--$18~\mathrm{ms}$ classical slowing-down estimates, these values indicate that charge exchange with background neutrals contributes appreciably to the measured decay. The conditioning dependence appears to be a robust experimental trend in these first data; however, comparison across identical discharges with varying lithium conditioning is not possible because changing wall conditions strongly affect plasma performance and shot evolution. To diagnose the balance between plasma heating and fueling as a function of lithium conditioning a forward model will simulataneously need to fit fast ion evolution and slowing down on electrons. The multi-modal array can constrain such future models using \lya{} for neutral-density profiles, AXUV signals for beam capture and fast-neutral loss, and soft-x-ray measurements with Thomson scattering for beam heating and plasma response.

\add{Disentangling the photon and direct fast-neutral contributions to the AXUV signal will require a coupled forward model. DEGAS2 can use the \lya{} measurements to constrain the background neutral distribution and synthetic \lya{} emissivity, as previously demonstrated on \ltx{} \cite{Maan_2024_recycling,DEGAS2}. NUBEAM can evolve beam deposition, fast-ion slowing down, charge-exchange loss and recapture \cite{Pankin2004NUBEAM}, while prompt orbit loss and beam-capture geometry can be constrained using the POET, CONBEAM, and 3dOrb calculations already applied to \ltx{} \cite{Capecchi_2021,Zakharov2026LTXBetaReport}. Folding the modeled photon and escaping-neutral fluxes through the corresponding detector responses and comparing them with the nearly overlapping \lya{} and AXUV measurements would provide a route to estimating the two AXUV contributions and testing how lithium conditioning modifies beam capture and charge-exchange loss \cite{Clary_2012,Veshchev_2006,Zhu_2012,Liu_2016}.}

\add{Future implementations could separate these contributions experimentally using nearly coincident AXUV channels with different front membranes, extending SSNPA designs that use overlapping diode views and calibrated tungsten coatings to suppress photon backgrounds and impose known neutral-energy cutoffs \cite{Zhu_2012,Liu_2014_SSNPA,Liu_2016}. A matched channel covered by a membrane selected to stop 12--20~keV H neutrals would provide a photon reference over its calibrated passband; after correcting for membrane transmission and relative etendue, its response could be subtracted from an unmasked channel to constrain the direct fast-neutral contribution \cite{Ziegler_2010,Henke1993}.}


\begin{acknowledgments}
This work was supported by the U.S. Department of Energy under contract number DE-AC02-09CH11466. The United States Government retains a non-exclusive, paid-up, irrevocable, world-wide license to publish or reproduce the published form of this manuscript, or allow others to do so, for United States Government purposes. 
\end{acknowledgments}

\section*{Data Availability}
The raw data used to produce the results in this manuscript can be accessed through the Princeton Data Commons at doi:10.34770/ehj9-ns15 \cite{tor_rsi_data}.

\bibliography{references}

@article{Hughes_2021,
  author =        {P E Hughes and W Capecchi and D B Elliott and
                   L E Zakharov and others},
  journal =       {Plasma Physics and Controlled Fusion},
  month =         {jul},
  number =        {8},
  pages =         {085020},
  publisher =     {IOP Publishing},
  title =         {Toroidal plasma acceleration due to NBI fast ion
                   losses in LTX-β},
  volume =        {63},
  year =          {2021},
  doi =           {10.1088/1361-6587/ac0b9f},
  url =           {https://dx.doi.org/10.1088/1361-6587/ac0b9f},
}

@article{Capecchi_2021,
  author =        {W. Capecchi and J.K. Anderson and D.P. Boyle and
                   P.E. Hughes and others},
  journal =       {Nuclear Fusion},
  month =         {oct},
  number =        {12},
  pages =         {126014},
  publisher =     {IOP Publishing},
  title =         {Neutral beam prompt loss in LTX-β},
  volume =        {61},
  year =          {2021},
  doi =           {10.1088/1741-4326/ac2bbf},
  url =           {https://dx.doi.org/10.1088/1741-4326/ac2bbf},
}

@article{McCracken_1979,
  author =        {G.M. McCracken and P.E. Stott},
  journal =       {Nuclear Fusion},
  month =         {jul},
  number =        {7},
  pages =         {889},
  publisher =     {},
  title =         {Plasma-surface interactions in tokamaks},
  volume =        {19},
  year =          {1979},
  doi =           {10.1088/0029-5515/19/7/004},
  url =           {https://dx.doi.org/10.1088/0029-5515/19/7/004},
}

@book{stangeby,
  author =        {C Stangeby, P},
  booktitle =     {The Plasma Boundary of Magnetic Fusion Devices.
                   Series: Series in Plasma Physics, ISBN:
                   978-0-7503-0559-4. Taylor \& Francis, Edited by Peter
                   Stangeby, vol. 7},
  month =         {01},
  pages =         {},
  publisher =     {Taylor \& Francis},
  title =         {{The Plasma Boundary of Manegic Fusion Devices}},
  year =          {2002},
}

@article{TSUJI1991311,
  author =        {Shunji Tsuji},
  journal =       {Fusion Engineering and Design},
  number =        {4},
  pages =         {311-324},
  title =         {Experimental scaling of particle confinement in
                   tokamaks},
  volume =        {15},
  year =          {1991},
  doi =           {https://doi.org/10.1016/0920-3796(92)90017-X},
  issn =          {0920-3796},
  url =           {https://www.sciencedirect.com/science/article/pii/
                  092037969290017X},
}

@misc{tor_rsi_data,
  author =        {Maan, Anurag},
  publisher =     {{Princeton Plasma Physics Laboratory, Princeton
                   University}},
  title =         {{Data for - Photodiode based multi-modal diagnostic
                   for low-energy neutral beam injection in the
                   LTX-$\beta$ spherical tokamak}},
  year =          {2026},
  url =           {https://doi.org/10.34770/ehj9-ns15},
}

@article{Baldwin_2002,
  author =        {M.J. Baldwin and R.P. Doerner and S.C. Luckhardt and
                   R.W. Conn},
  journal =       {Nuclear Fusion},
  month =         {oct},
  number =        {11},
  pages =         {1318},
  publisher =     {},
  title =         {Deuterium retention in liquid lithium},
  volume =        {42},
  year =          {2002},
  doi =           {10.1088/0029-5515/42/11/305},
  url =           {https://dx.doi.org/10.1088/0029-5515/42/11/305},
}

@article{Kaita_2019,
  author =        {Robert Kaita},
  journal =       {Plasma Physics and Controlled Fusion},
  month =         {oct},
  number =        {11},
  pages =         {113001},
  publisher =     {IOP Publishing},
  title =         {Fusion applications for lithium: wall conditioning in
                   magnetic confinement devices},
  volume =        {61},
  year =          {2019},
  doi =           {10.1088/1361-6587/ab4156},
  url =           {https://dx.doi.org/10.1088/1361-6587/ab4156},
}

@article{crashninicove,
  author =        {Krasheninnikov, S. I. and Zakharov, L. E. and
                   Pereverzev, G. V.},
  journal =       {Physics of Plasmas},
  month =         {04},
  number =        {5},
  pages =         {1678-1682},
  title =         {{On lithium walls and the performance of magnetic
                   fusion devices}},
  volume =        {10},
  year =          {2003},
  doi =           {10.1063/1.1558293},
  issn =          {1070-664X},
  url =           {https://doi.org/10.1063/1.1558293},
}

@article{isomak,
  author =        {Catto, Peter J. and Hazeltine, R. D.},
  journal =       {Physics of Plasmas},
  month =         {12},
  number =        {12},
  pages =         {122508},
  title =         {{Isothermal tokamak}},
  volume =        {13},
  year =          {2006},
  doi =           {10.1063/1.2403090},
  issn =          {1070-664X},
  url =           {https://doi.org/10.1063/1.2403090},
}

@article{boyle-prl,
  author =        {Boyle, D. P. and Majeski, R. and Schmitt, J. C. and
                   Hansen, C. and Kaita, R. and Kubota, S. and Lucia, M. and
                   Rognlien, T. D.},
  journal =       {Phys. Rev. Lett.},
  month =         {Jul},
  pages =         {015001},
  publisher =     {American Physical Society},
  title =         {{Observation of Flat Electron Temperature Profiles in
                   the Lithium Tokamak Experiment}},
  volume =        {119},
  year =          {2017},
  doi =           {10.1103/PhysRevLett.119.015001},
  url =           {https://link.aps.org/doi/10.1103/PhysRevLett.119.015001},
}

@article{majeski_pop,
  author =        {Majeski, R. and Bell, R. E. and Boyle, D. P. and
                   Kaita, R. and others},
  journal =       {Physics of Plasmas},
  month =         {03},
  number =        {5},
  pages =         {056110},
  title =         {{Compatibility of lithium plasma-facing surfaces with
                   high edge temperatures in the Lithium Tokamak
                   Experiment}},
  volume =        {24},
  year =          {2017},
  doi =           {10.1063/1.4977916},
  issn =          {1070-664X},
  url =           {https://doi.org/10.1063/1.4977916},
}

@article{MAAN2023_NME,
  author =        {A. Maan and D.P. Boyle and R. Majeski and S. Banerjee and
                   others},
  journal =       {Nuclear Materials and Energy},
  pages =         {101408},
  title =         {Improved neutral and plasma density control with
                   increasing lithium wall coatings in the Lithium
                   Tokamak Experiment-β (LTX-β)},
  volume =        {35},
  year =          {2023},
  doi =           {https://doi.org/10.1016/j.nme.2023.101408},
  issn =          {2352-1791},
  url =           {https://www.sciencedirect.com/science/article/pii/
                  S2352179123000479},
}

@article{Maan_2024_recycling,
  author =        {A. Maan and D. P. Boyle and R. Majeski and
                   G. J. Wilkie and others},
  journal =       {Physics of Plasmas},
  number =        {2},
  pages =         {022505},
  title =         {Estimates of global recycling coefficients for
                   {LTX-$\beta$} discharges},
  volume =        {31},
  year =          {2024},
  doi =           {10.1063/5.0177604},
  url =           {https://doi.org/10.1063/5.0177604},
}

@article{Boyle_NF_2023,
  author =        {D.P. Boyle and J. Anderson and S. Banerjee and
                   R.E. Bell and others},
  journal =       {Nuclear Fusion},
  month =         {apr},
  number =        {5},
  pages =         {056020},
  publisher =     {IOP Publishing},
  title =         {Extending the low-recycling, flat temperature profile
                   regime in the lithium tokamak experiment-β (LTX-β)
                   with ohmic and neutral beam heating},
  volume =        {63},
  year =          {2023},
  doi =           {10.1088/1741-4326/acc4da},
  url =           {https://dx.doi.org/10.1088/1741-4326/acc4da},
}

@article{Clary_2012,
  author =        {Clary, R. and Smirnov, A. and Dettrick, S. and
                   Knapp, K. and others},
  journal =       {Review of Scientific Instruments},
  number =        {10},
  pages =         {10D713},
  title =         {A photodiode-based neutral particle bolometer for
                   characterizing charge-exchanged fast-ion behavior},
  volume =        {83},
  year =          {2012},
  doi =           {10.1063/1.4732860},
  url =           {https://doi.org/10.1063/1.4732860},
}

@article{Veshchev_2006,
  author =        {Veshchev, E. A. and Ozaki, T. and Goncharov, P. R. and
                   Sudo, S. and LHD Experimental Group},
  journal =       {Review of Scientific Instruments},
  number =        {10},
  pages =         {10F129},
  title =         {Initial angle resolved measurements of fast neutrals
                   using a multichannel linear {AXUV} detector system on
                   {LHD}},
  volume =        {77},
  year =          {2006},
  doi =           {10.1063/1.2351916},
  url =           {https://doi.org/10.1063/1.2351916},
}

@article{Zhu_2012,
  author =        {Zhu, Y. B. and Bortolon, A. and Heidbrink, W. W. and
                   Celle, S. L. and Roquemore, A. L.},
  journal =       {Review of Scientific Instruments},
  number =        {10},
  pages =         {10D304},
  title =         {Compact solid-state neutral particle analyzer in
                   current mode},
  volume =        {83},
  year =          {2012},
  doi =           {10.1063/1.4732070},
  url =           {https://doi.org/10.1063/1.4732070},
}

@article{Liu_2016,
  author =        {Liu, D. and Heidbrink, W. W. and Tritz, K. and
                   others},
  journal =       {Review of Scientific Instruments},
  number =        {11},
  pages =         {11D803},
  title =         {Compact and multi-view solid state neutral particle
                   analyzer arrays on {National Spherical Torus
                   Experiment-Upgrade}},
  volume =        {87},
  year =          {2016},
  doi =           {10.1063/1.4959798},
  url =           {https://doi.org/10.1063/1.4959798},
}

@article{Degeling_2004,
  author =        {Degeling, A. W. and Weisen, H. and Zabolotsky, A. and
                   Duval, B. P. and Pitts, R. A. and Wischmeier, M. and
                   Lavanchy, P. and Marmillod, Ph. and Pochon, G.},
  journal =       {Review of Scientific Instruments},
  number =        {10},
  pages =         {4139--4141},
  title =         {{AXUV} bolometer and {Lyman-alpha} camera systems on
                   the {TCV} tokamak},
  volume =        {75},
  year =          {2004},
  doi =           {10.1063/1.1787131},
  url =           {https://doi.org/10.1063/1.1787131},
}

@article{lama_cal,
  author =        {Laggner, F. M. and Bortolon, A. and Rosenthal, A. M. and
                   Wilks, T. M. and Hughes, J. W. and Freeman, C. and
                   Golfinopoulos, T. and Nagy, A. and Mauzey, D. and
                   Shafer, M. W. and the DIII-D Team},
  journal =       {Review of Scientific Instruments},
  month =         {03},
  number =        {3},
  pages =         {033522},
  title =         {{Absolute calibration of the Lyman-α measurement
                   apparatus at DIII-D}},
  volume =        {92},
  year =          {2021},
  doi =           {10.1063/5.0038134},
  issn =          {0034-6748},
  url =           {https://doi.org/10.1063/5.0038134},
}

@article{Krstic_2023_LiPMI,
  author =        {P. S. Krstic and E. T. Ostrowski and S. Dwivedi and
                   S. Abe and A. Maan and A. C. T. van Duin and
                   B. E. Koel},
  journal =       {Journal of Applied Physics},
  month =         {sep},
  number =        {10},
  pages =         {100902},
  title =         {Detailed studies of the processes in low energy H
                   irradiation of Li and Li-compound surfaces},
  volume =        {134},
  year =          {2023},
  doi =           {10.1063/5.0149502},
  url =           {https://doi.org/10.1063/5.0149502},
}

@article{tosh,
  author =        {Le, Tosh Xavier Keating and Banerjee, Santanu and
                   Maan, Anurag and Majeski, Richard P and
                   Boyle, Dennis P and Shousha, Ricardo and Li, Boting and
                   Morales, Javier and López Pérez, Camila and
                   Gajani, Hussain and Kubota, Shigeyuki},
  journal =       {Plasma Physics and Controlled Fusion},
  title =         {Synthetic Modeling of Soft X-Ray Emissivity for
                   Magnetic Island Analysis in LTX-β},
  year =          {2026},
  url =           {http://iopscience.iop.org/article/10.1088/1361-6587/ae6d6f},
}

@article{Banerjee_2024,
  author =        {Banerjee, Santanu and Boyle, D.P. and Maan, A. and
                   Ferraro, N. and Wilkie, G. and Majeski, R. and
                   Podesta, M. and Bell, R. and Hansen, C. and
                   Capecchi, W. and Elliott, D.},
  journal =       {Nuclear Fusion},
  month =         {mar},
  number =        {4},
  pages =         {046026},
  publisher =     {IOP Publishing},
  title =         {Investigating the role of edge neutrals in exciting
                   tearing mode activity and achieving flat temperature
                   profiles in LTX-β},
  volume =        {64},
  year =          {2024},
  doi =           {10.1088/1741-4326/ad2ca7},
  url =           {https://doi.org/10.1088/1741-4326/ad2ca7},
}

@misc{OptoDiodeAXUVSeries,
  author =        {{Opto Diode Corporation}},
  howpublished =  {\url{https://optodiode.com/product-category/axuv/}},
  note =          {Accessed August 7, 2026},
  title =         {{AXUV} Photodiodes: High-Energy Photon, Electron, and
                   X-Ray Detection},
  year =          {2026},
}

@article{Henke1993,
  author =        {Henke, B. L. and Gullikson, E. M. and Davis, J. C.},
  journal =       {Atomic Data and Nuclear Data Tables},
  number =        {2},
  pages =         {181--342},
  title =         {X-ray interactions: Photoabsorption, scattering,
                   transmission, and reflection at $E=50$--30000 eV,
                   $Z=1$--92},
  volume =        {54},
  year =          {1993},
  doi =           {10.1006/adnd.1993.1013},
}

@misc{CXROFilterTransmission,
  author =        {{Center for X-Ray Optics}},
  howpublished =
  {\url{https://henke.lbl.gov/optical_constants/filter2.html}},
  note =          {Beryllium filter calculation using density
                   $1.848~\mathrm{g\,cm^{-3}}$ and thickness
                   $5~\mu\mathrm{m}$},
  title =         {Filter Transmission Calculator},
  year =          {2026},
}

@techreport{Zakharov2026LTXBetaReport,
  author =        {Zakharov, L. E.},
  institution =   {LiWFusion},
  month =         {feb},
  note =          {Final report for Award Number DE-SC0023274},
  number =        {RPT-0000018431},
  title =         {Understanding {NBI} heating and fueling in
                   {LTX-$\beta$} tokamak},
  year =          {2026},
}

@article{Hogan1997,
  author =        {Hogan, J. T. and Maingi, R. and Mioduszewski, P. K. and
                   Hutter, Th. and Klepper, C. C. and Wade, M. R.},
  journal =       {Journal of Nuclear Materials},
  pages =         {612--617},
  title =         {Core/divertor/wall particle dynamics in the {DIII-D}
                   tokamak},
  volume =        {241--243},
  year =          {1997},
  doi =           {10.1016/S0022-3115(97)80109-2},
  url =           {https://doi.org/10.1016/S0022-3115(97)80109-2},
}

@article{Bagryansky1999,
  author =        {Bagryansky, P. A. and Bender, E. D. and Ivanov, A. A. and
                   Karpushov, A. N. and Murachtin, S. V. and Noack, K. and
                   Krahl, St. and Collatz, S.},
  journal =       {Journal of Nuclear Materials},
  number =        {1--2},
  pages =         {124--133},
  title =         {Effect of fast {Ti}-deposition on gas recycling at
                   the first wall and on fast ion losses in the {GDT}
                   experiment},
  volume =        {265},
  year =          {1999},
  doi =           {10.1016/S0022-3115(98)00510-8},
  url =           {https://doi.org/10.1016/S0022-3115(98)00510-8},
}

@article{Lucia_2015_surface,
  author =        {Lucia, M. and Kaita, R. and Majeski, R. and
                   Bedoya, F. and others},
  journal =       {Journal of Nuclear Materials},
  pages =         {907--910},
  title =         {Dependence of {LTX} plasma performance on surface
                   conditions as determined by in situ analysis of
                   plasma facing components},
  volume =        {463},
  year =          {2015},
  doi =           {10.1016/j.jnucmat.2014.11.006},
  url =           {https://doi.org/10.1016/j.jnucmat.2014.11.006},
}

@article{Maan_PPCF_2020,
  author =        {A Maan and D P Boyle and R Kaita and E T Ostrowski and
                   others},
  journal =       {Plasma Physics and Controlled Fusion},
  month =         {dec},
  number =        {2},
  pages =         {025007},
  publisher =     {{IOP} Publishing},
  title =         {Oxidation of lithium plasma facing components and its
                   effect on plasma performance in the lithium tokamak
                   experiment-$\beta$},
  volume =        {63},
  year =          {2020},
  doi =           {10.1088/1361-6587/abcd0f},
  url =           {https://doi.org/10.1088/1361-6587/abcd0f},
}

@article{Maan_IEEE_2020,
  author =        {Maan, Anurag and Ostrowski, Evan and Kaita, Robert and
                   Donovan, David and others},
  journal =       {IEEE Transactions on Plasma Science},
  number =        {6},
  pages =         {1463-1467},
  title =         {Plasma Facing Component Characterization and
                   Correlation With Plasma Conditions in Lithium Tokamak
                   Experiment-$\beta$},
  volume =        {48},
  year =          {2020},
  doi =           {10.1109/TPS.2020.2969115},
}

@article{Elliott_2018_CHERS,
  author =        {Elliott, D. B. and Biewer, T. M. and Boyle, D. P. and
                   Kaita, R. and Majeski, R.},
  journal =       {Review of Scientific Instruments},
  month =         {10},
  number =        {10},
  pages =         {10D118},
  title =         {{The charge exchange recombination spectroscopy
                   diagnostic on the upgraded Lithium Tokamak Experiment
                   (LTX-$\beta$)}},
  volume =        {89},
  year =          {2018},
  doi =           {10.1063/1.5039368},
  url =           {https://doi.org/10.1063/1.5039368},
}

@misc{DEGAS2,
  author =        {D. Stotler and C. Karney and R. Kanzleiter and
                   S. Jaishankar},
  howpublished =  {\url{https://w3.pppl.gov/degas2/} (Sep, 23, 2020)},
  title =         {DEGAS 2 User's Manual},
  year =          {2020},
}

@techreport{Barnett1990AtomicData,
  author =        {Barnett, C. F. and Hunter, H. T. and
                   Kirkpatrick, M. I. and Alvarez, I. and Cisneros, C. and
                   Phaneuf, R. A.},
  institution =   {Oak Ridge National Laboratory},
  month =         {jul},
  number =        {ORNL-6086/V1},
  title =         {Atomic Data for Fusion, Volume 1: Collisions of {H},
                   {H2}, {He} and {Li} Atoms and Ions with Atoms and
                   Molecules},
  year =          {1990},
  doi =           {10.2172/6570226},
  url =           {https://doi.org/10.2172/6570226},
}

@misc{Maan_2026_fLITER,
  author =        {A. Maan and R. Majeski and C. L{\'o}pez P{\'e}rez and
                   D. P. Boyle and T. Le and R. Lunsford},
  title =         {Development of flash lithium evaporators for
                   {NSTX-U}},
  year =          {2026},
  doi =           {10.48550/arXiv.2606.00777},
  url =           {https://arxiv.org/abs/2606.00777},
}

@article{Boyle_NME_2025,
  author =        {Boyle, D. P. and Maan, A. and Majeski, R. and
                   Ostrowski, E. and Abe, S. and Banerjee, S. and
                   Capecchi, W. and Elliott, D. B. and Hansen, C. and
                   Jung, E. and Hughes, P. E. and Kubota, S. and
                   Lampert, M. and Maingi, R. and McLean, A. G. and
                   Menard, J. E. and Morales, J. J. and
                   Soukhanovskii, V. A.},
  journal =       {Nuclear Materials and Energy},
  pages =         {101949},
  title =         {Improved liquid lithium surfaces in the Lithium
                   Tokamak Experiment-$\beta$},
  volume =        {43},
  year =          {2025},
  doi =           {10.1016/j.nme.2025.101949},
  url =           {https://doi.org/10.1016/j.nme.2025.101949},
}

@inproceedings{Summers2007ADAS,
  address =       {Melville, NY},
  author =        {Summers, H. P. and O'Mullane, M. G. and
                   Whiteford, A. D. and Badnell, N. R. and Loch, S. D.},
  booktitle =     {Atomic and Molecular Data and Their Applications},
  editor =        {Roueff, Evelyne},
  pages =         {239--248},
  publisher =     {American Institute of Physics},
  series =        {AIP Conference Proceedings},
  title =         {{ADAS}: Atomic data, modelling and analysis for
                   fusion},
  volume =        {901},
  year =          {2007},
  isbn =          {0735404070},
  url =           {https://pureportal.strath.ac.uk/en/publications/adas-atomic-
                  data-modelling-and-analysis-for-fusion/},
}

@misc{ADAS302ADF02,
  author =        {{ADAS Project}},
  howpublished =  {{ADAS} User Manual, Chapter 3-02},
  month =         {mar},
  note =          {{ADF02} ion-atom reaction cross-section collections},
  title =         {{ADAS302}: Ion/atom data -- graph and fit
                   cross-section},
  year =          {2003},
  url =           {https://www.adas.ac.uk/man/chap3-02.pdf},
}

@article{Pankin2004NUBEAM,
  author =        {Pankin, A. and McCune, D. and Andre, R. and
                   Bateman, G. and Kritz, A.},
  journal =       {Computer Physics Communications},
  number =        {3},
  pages =         {157--184},
  title =         {The tokamak {Monte Carlo} fast ion module {NUBEAM} in
                   the {National Transport Code Collaboration} library},
  volume =        {159},
  year =          {2004},
  doi =           {10.1016/j.cpc.2003.11.002},
  url =           {https://doi.org/10.1016/j.cpc.2003.11.002},
}

@article{Liu_2014_SSNPA,
  author =        {Liu, D. and Heidbrink, W. W. and Tritz, K. and
                   Zhu, Y. B. and Roquemore, A. L. and Medley, S. S.},
  journal =       {Review of Scientific Instruments},
  number =        {11},
  pages =         {11E105},
  title =         {Design of solid state neutral particle analyzer array
                   for {National Spherical Torus Experiment-Upgrade}},
  volume =        {85},
  year =          {2014},
  doi =           {10.1063/1.4889913},
  url =           {https://doi.org/10.1063/1.4889913},
}

@article{Ziegler_2010,
  author =        {Ziegler, J. F. and Ziegler, M. D. and
                   Biersack, J. P.},
  journal =       {Nuclear Instruments and Methods in Physics Research
                   Section B: Beam Interactions with Materials and
                   Atoms},
  number =        {11--12},
  pages =         {1818--1823},
  title =         {{SRIM}---The stopping and range of ions in matter
                   (2010)},
  volume =        {268},
  year =          {2010},
  doi =           {10.1016/j.nimb.2010.02.091},
  url =           {https://doi.org/10.1016/j.nimb.2010.02.091},
}

\end{document}